# Comparative Analysis of Conventional and Modified H- Bridge Inverter Configuration

Chithaj Mallikarjun
Dept. Electrical and Electronics Engineering
PES University
Bengaluru, India
chithaj95@gmail.com

Niteesh S. Shanbog
Dept. Electrical and Electronics Engineering
PES University
Bengaluru, India
nit1995@yahoo.co.in

Sangeeta Modi
Dept. Electrical and Electronics Engineering
PES University
Bengaluru, India
smodi@pes.edu

*Abstract*— **In this paper a comparative analysis is presented for conventional and modified H Bridge configuration of 5 and 7 level inverter. A modified H Bridge converter utilizes a reduced number of switches for the same level output as compared to the conventional H-Bridge configuration. The lower number of switches will result in reduced switching losses, installation cost and converter cost. MATLAB/SIMULINK software is used for simulation of the different configurations used for the comparison. R and RL type of load is used and the corresponding voltage waveform is analyzed for its harmonic content. It can be seen from the results obtained that the Total Harmonic Distortion (THD) in the modified and conventional 7 level configuration is less than that of the 5 level inverter configurations.**

I. INTRODUCTION

The increase in demand of power supply has led to increased focus on the development of renewable energy sources. Solar energy is one such major source of renewable energy. Solar energy produces power which is dc in nature. This dc power will then have to be converted into ac for in order to make it compatible with the present day equipments. This conversion of dc power to ac is done using cascaded H bridge multilevel inverters with preferably low THD[1]. This study has the potential to help a design engineer to select an appropriate multilevel inverter for the application required. Multilevel inverters can be classified as current source inverters or voltage source inverters. In the case of a multilevel current source inverter, a short circuit in the circuit can cause very fault current which will damage any other equipments connected to the circuit. Hence, multilevel voltage source inverters preferred [2]. Multilevel voltage source inverters can be classified into three main categories as (i) Cascaded H- bridge multilevel inverter, (ii) Neutral point clamped multilevel inverter and (iii) Flying capacitor multilevel inverter. Cascaded H bridge multilevel inverter is more preferred because it gives power levels, high output voltage, and reliability. The control of the switches in such an inverter is also simple and easy to construct. The cascaded H-bridge inverters also use reduced number of switches and have a modular structure which makes it easy to fabricate. Drive applications which have high power and high voltage ratings also make use of these inverters. High power and power quality applications also make use of these H-bridge multilevel inverters. Some of the other applications are photovoltaic power conversion, reactive power compensation applications, magnetic resonance imaging, static synchronous compensators active filter, magnetic resonance imaging uninterruptible power supplies, and magnetic resonance imaging. Potential applications include electric and hybrid power trains. Any distortion in the grid due to the harmonic content will have to be reduced in order to have maximum energy efficiency [3]. Lakshmi et al [1] introduced cascade seven level inverter which has reduced number of switches which operate using a level shifting PWM technique. A review on multilevel inverter configuration was proposed by Pharne et al [2]. A new multilevel converter configuration with reduced number of power electronic components was described by Ebrahimi et al [4]. A new configuration of Cascaded multilevel converters with reduced number of components for high-voltage applications was proposed by Babaei et al [6] . Another new multilevel inverter configuration's design and implementation was proposed by Najafi et al [5] . Similarly, a cascade multilevel converter configuration with reduced number of switches was described by Ebrahim Babaei[8]. A survey of multilevel inverter configurations, controls and applications was made by Rodriguez et al [9]. A cascaded multilevel inverter with regeneration capability and a reduced number of switches was dealt with by Lezana et al [7].

## II. CONVENTIONAL CASCADED MULTILEVEL INVERTER TOPOLOGY

Renewable energy sources which usually contain separate dc source like photovoltaic battery, fuel cell, ultra capacitor use cascaded multilevel inverter. The back-to-back fashion between two converters is not possible in this inverter. It requires lower number of components in order to get the same number of voltage level. There is no need of using extra diodes and capacitors. Because of the same structure, it is scalable. It is possible to have a modularized circuit plan, thus making it easier for fabrication. Power conversion systems use multilevel inverters because of its improved voltage and current waveforms. The performance and efficiency of the system is therefore improved. Multilevel inverters have specialized applications in high voltage AC motor drive, distributive generation, high voltage direct transmission as well as in SVC.

*A. Cascaded H-Bridge 5 level Inverter Configuration.*

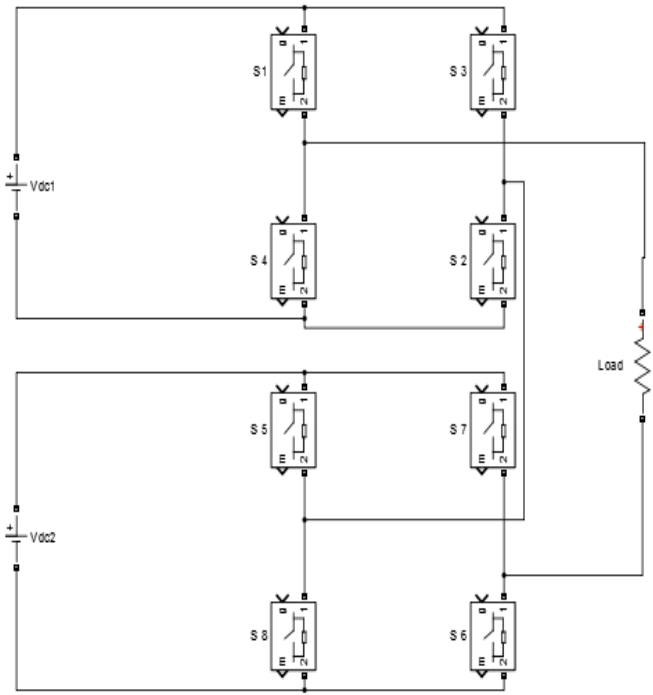

Fig.1   5 Level Conventional Cascaded H-bridge Inverter

The cascaded 5-level H-bridge configuration is shown in Fig 1. This inverter uses 2 separate DC sources and 8 switching devices. A five level output of 2Vdc, 1Vdc, 0, -1Vdc, -2Vdc is generated by this configuration. The switching table for the switches in the five level H-bridge configuration in Table 1. The R or RL load is connected between the terminals.

TABLE 1
SWITCHING STATES OF THE PROPOSED CONFIGURATION

| Modes | S1 | S2 | S3 | S4 | S5 | S6 | S7 | S8 |
|---|---|---|---|---|---|---|---|---|
| 2 Vdc | 1 | 1 | 0 | 0 | 1 | 1 | 0 | 0 |
| Vdc | 1 | 1 | 0 | 0 | 0 | 1 | 0 | 1 |
| 0 | 0 | 0 | 0 | 0 | 0 | 0 | 0 | 0 |
| - Vdc | 0 | 0 | 1 | 1 | 0 | 1 | 0 | 1 |
| -2 Vdc | 0 | 0 | 1 | 1 | 0 | 0 | 1 | 1 |

*B. Cascaded 7 level H-Bridge Inverter Configuration.*

The cascaded 7-level H-bridge configuration is shown in Fig 2. This inverter uses 3 separate DC sources and 12 switching devices. A seven level output of 3Vdc, 2Vdc, 1Vdc, 0, -1Vdc, -2Vdc, -3Vdc is generated by this configuration. The switching table for the switches in the seven level H-bridge configuration in Table 2. The R or RL load is connected between the terminals.

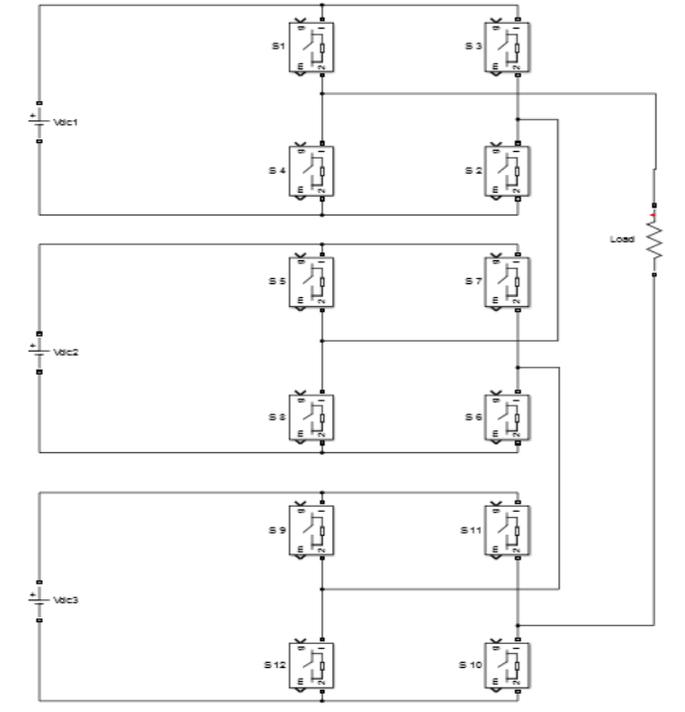

Fig.2   7 Level Conventional Cascaded H-bridge Inverter

TABLE 2
SWITCHING STATES OF THE PROPOSED CONFIGURATION

| Modes | S1 | S2 | S3 | S4 | S5 | S6 | S7 | S8 | S9 | S10 | S11 | S12 |
|---|---|---|---|---|---|---|---|---|---|---|---|---|
| 3 Vdc | 1 | 1 | 0 | 0 | 1 | 1 | 0 | 0 | 1 | 1 | 0 | 0 |
| 2 Vdc | 1 | 1 | 0 | 0 | 1 | 1 | 0 | 0 | 0 | 1 | 0 | 1 |
| Vdc | 1 | 1 | 0 | 0 | 0 | 1 | 0 | 1 | 0 | 1 | 0 | 1 |
| 0 | 0 | 0 | 0 | 0 | 0 | 0 | 0 | 0 | 0 | 0 | 0 | 0 |
| - Vdc | 0 | 0 | 1 | 1 | 0 | 1 | 0 | 1 | 0 | 1 | 0 | 1 |
| -2 Vdc | 0 | 0 | 1 | 1 | 0 | 0 | 1 | 1 | 0 | 1 | 0 | 1 |
| -3 Vdc | 0 | 0 | 1 | 1 | 0 | 0 | 1 | 1 | 0 | 0 | 1 | 1 |

## III. MODIFIED CONFIGURATION

The multilevel inverter is mainly used to improve the quality of output voltage. It also reduces the number of switches. It is important to obtain a sinusoidal output voltage waveform with lower order harmonics. The main concern of the fundamental switching scheme is to obtain switching angles in order to produce a voltage with fundamental frequency. For increasing voltage levels, the number of switches required will also increase. This increases switching losses, voltage stresses, and also the complexity of the circuit. In the modified configuration, the efficiency of the system is increased by reducing the number of switches. The lower order harmonics cause more effect on the average voltage as compared to the harmonics of higher order. The effect of third order harmonics is especially damaging on a motor load. This paper proposes method to eliminate lower order harmonics. In the H-bridge when the switches in the same leg conduct simultaneously, it leads to a short circuit condition. The harmonic content can be reduced by using techniques like Level shifting technique and Phase shifting technique. In this study level shifting technique is used to eliminate lower order harmonics. This also reduces the THD significantly. The modified system is also analysed with either R or R-L load.

### A. Modified 5 level Inverter Configuration

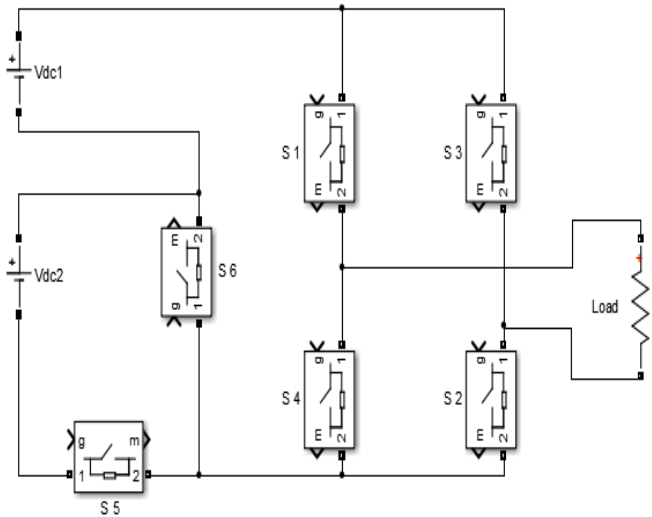

Fig.3  Modified 5 level Inverter Configuration.

Fig. 3 shows the cascaded 5-level proposed configuration. This configuration consists of 2 separate DC sources and 6 switching devices. It produces seven level output of 2Vdc, 1Vdc, 0, -1Vdc, -2Vdc. The switching table for the switches in the five level modified configuration is given in Table 3. The R or RL load is connected between terminals.

TABLE 3
SWITCHING STATES FOR THE MODIFIED 5 LEVEL CONFIGURATION

| Modes | S1 | S2 | S3 | S4 | S5 | S6 |
|---|---|---|---|---|---|---|
| 2 Vdc | 1 | 1 | 0 | 0 | 1 | 0 |
| Vdc | 1 | 1 | 0 | 0 | 0 | 1 |
| 0 | 0 | 0 | 0 | 0 | 0 | 0 |
| - Vdc | 0 | 0 | 1 | 1 | 0 | 1 |
| -2 Vdc | 0 | 0 | 1 | 1 | 1 | 0 |

### B. Modified 7 level Inverter Configuration

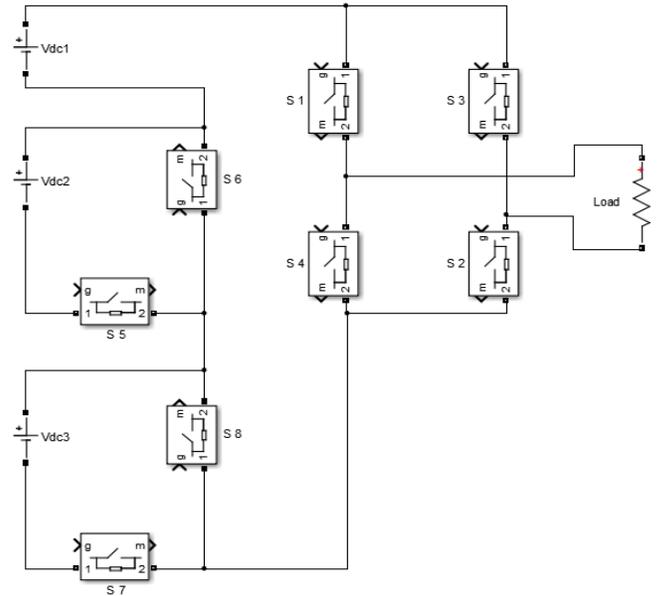

Fig.4  Modified 7 level Inverter Configuration.

Fig. 4 shows the cascaded 7-level proposed configuration. This configuration consists of 3 separate DC sources and 8 switching devices. A seven level output of 3Vdc, 2Vdc, 1Vdc, 0, -1Vdc, -2Vdc, -3Vdc is generated. The switching table for the switches in the seven level proposed configuration is given in Table 4. The R or RL load is connected between terminals.

TABLE 4
SWITCHING STATES FOR THE MODIFIED 7 LEVEL CONFIGURATION

| Modes | S1 | S2 | S3 | S4 | S5 | S6 | S7 | S8 |
|---|---|---|---|---|---|---|---|---|
| 3 Vdc | 1 | 1 | 0 | 0 | 1 | 0 | 1 | 0 |
| 2 Vdc | 1 | 1 | 0 | 0 | 1 | 0 | 0 | 1 |
| Vdc | 1 | 1 | 0 | 0 | 0 | 1 | 0 | 1 |
| 0 | 0 | 0 | 0 | 0 | 0 | 0 | 0 | 0 |
| - Vdc | 0 | 0 | 1 | 1 | 0 | 1 | 0 | 1 |
| -2 Vdc | 0 | 0 | 1 | 1 | 1 | 0 | 0 | 1 |
| -3 Vdc | 0 | 0 | 1 | 1 | 1 | 0 | 1 | 0 |

# IV SIMULATION RESULTS

Multilevel inverters of two levels (five and seven) are chosen for the purpose of comparison in this paper. Both conventional and the modified topologies are simulated in order to better gauge their performance with an R and RL load.

The figures below shows the simulated model. Each of the switches requires a separate gate driver circuits when used in practical circuits. A 100Volt DC power source is used along with a resistance 10ohms and an inductance of 100uH. MATLAB Simulink was used to analyze the harmonic spectrum present in the output voltage.

Total harmonic distortion, or THD, is defined as the total sum of all the harmonic components of the voltage or the current waveform compared to the fundamental component of the voltage or current wave:

$$V_1 = \frac{\sqrt{(V_2^2 + V_3^2 + V_4^2 + \cdots \ldots \ldots V_n^2)}}{V_1} * 100$$

The above formula is used for calculating the THD on a voltage signal. In this formula, the square root of the sum of the square of all the harmonic components is added which is then divided with its fundamental component to get the THD. Higher THD means more distortion in the main signal. Using voltages with low THD values, will ensure proper operation of equipment and increase the life span of the equipment.

A. 5 Level Inverter

Two types of 5 level inverters are designed and studied with R and RL load. The total harmonic distortion in the output wave is used to compare the harmonic content in both the topologies.

*1) 5 Level Conventional H Bridge inverter configuration:* The five level conventional inverter topology with 8 switches is simulated.

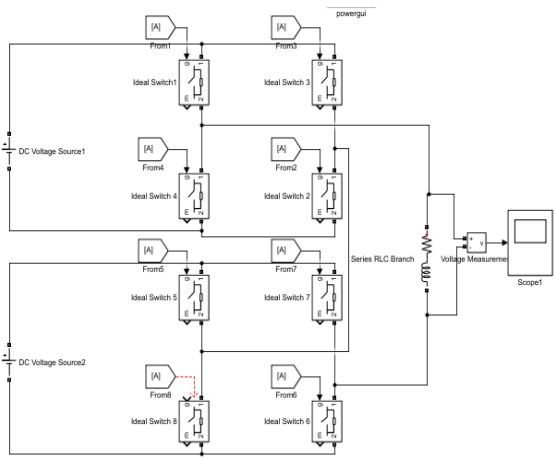

Fig. 5. Conventional 5 level inverter configuration

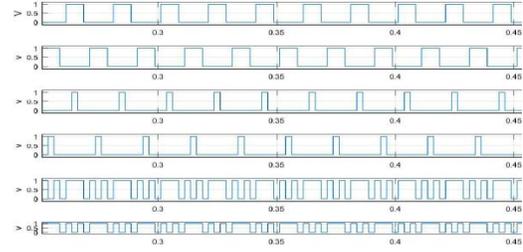

Fig. 6. Control signals for conventional 5 level inverter

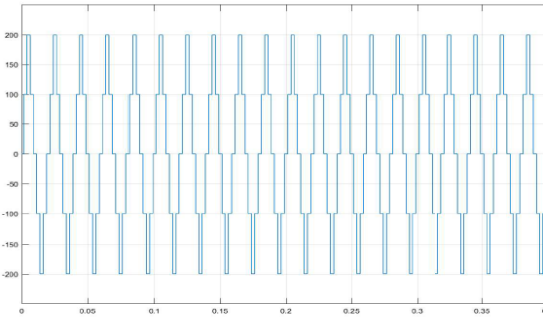

Fig. 7. Output voltage for conventional 5 level inverter with RL load

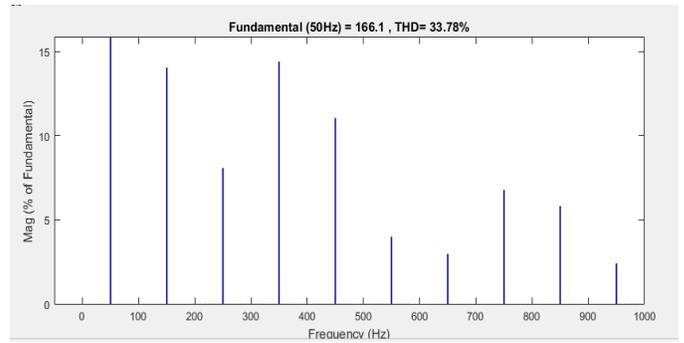

Fig. 8. THD values for the output voltage of a 5 level conventional inverter with RL load

*2) Modified 5 level inverter configuration:* The five level modified inverter with 6 switches is simulated using MATLAB Simulink.

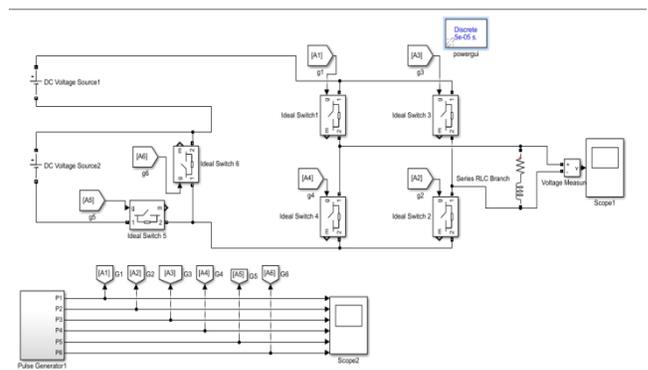

Fig. 9. Modified 5 level inverter configuration

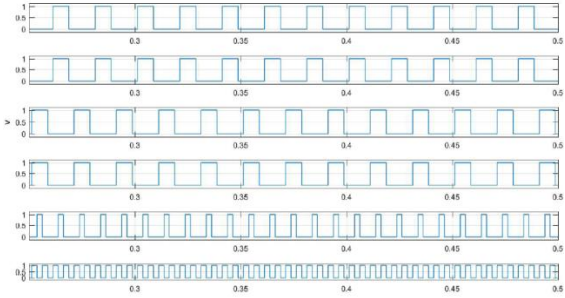

Fig. 10. Control signals for modified 5 level inverter

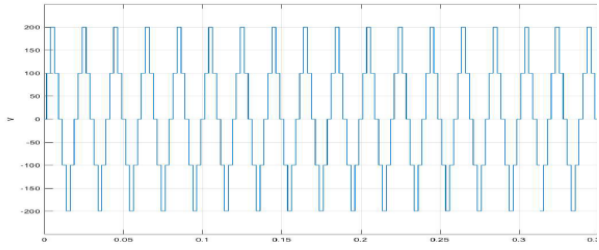

Fig. 11. Output voltage for modified 5 level inverter with RL load

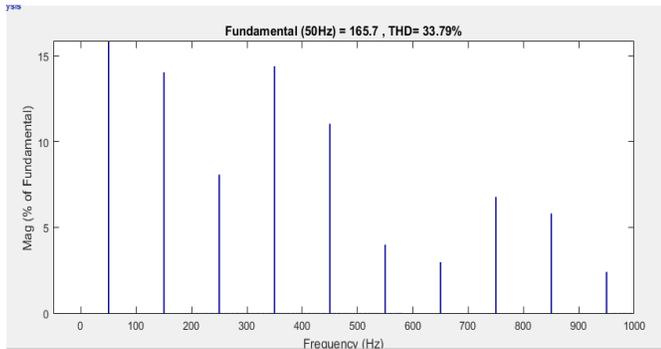

Fig. 12. THD values for the output voltage of a modified 5 level inverter with RL load

## C. Conventional 7 Level H Bridge Inverter

Two types of 7 level inverters are designed and studied with R and RL load. The harmonic content in both the topologies are compared using the total harmonic distortion in the output wave.

*1) Conventional 7 level inverter topology:* The conventional seven level inverter topology with 12 switches is simulated using MATLAB Simulink.

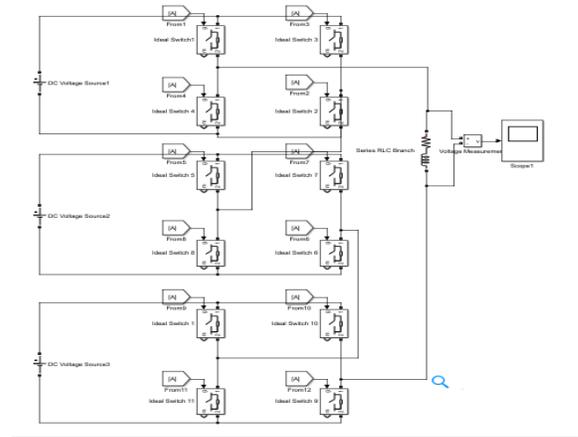

Fig. 13. Conventional 7 level inverter configuration

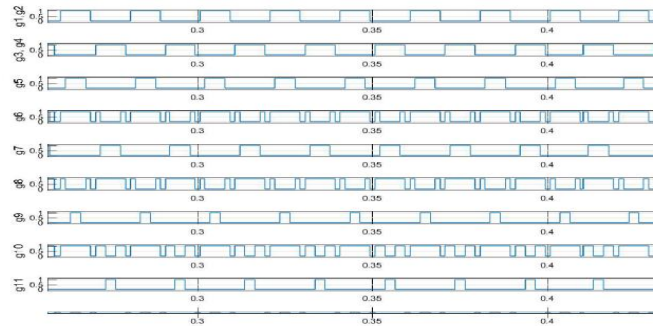

Fig. 14. Control signals for the conventional 7 level inverter

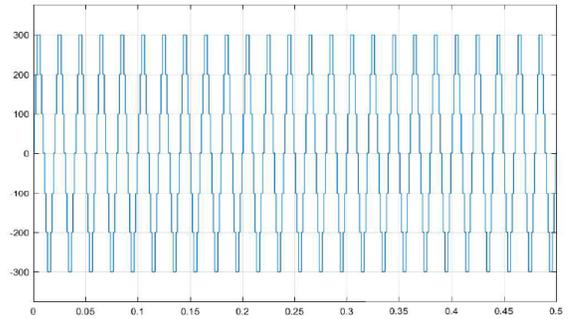

Fig. 15. Output voltage for conventional 7 level inverter with RL load

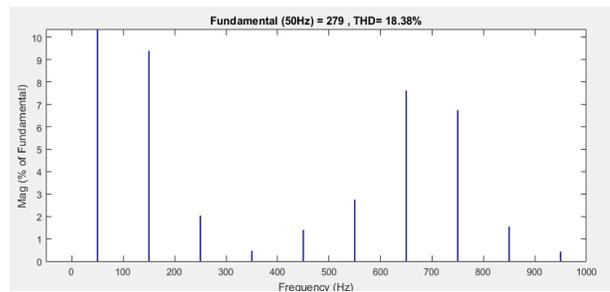

Fig. 16. THD values for the output voltage of a 7 level conventional inverter with RL load

*2) Modified 7 level inverter topology:* The seven level modified inverter with 12 switches is simulated using MATLAB Simulink.

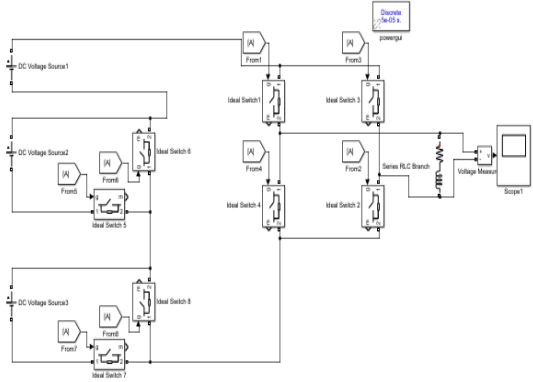

Fig. 17. Modified 7 level inverter configuration

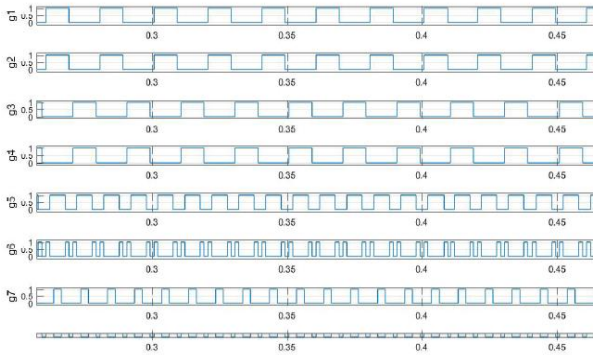

Fig. 18. Control signals for modified 7 level inverter configuration

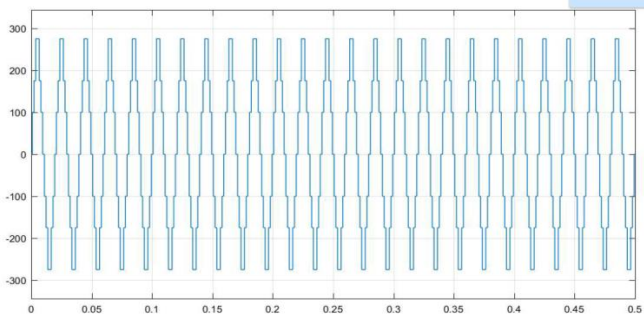

Fig. 19. Output voltage for modified 7 level inverter with RL load

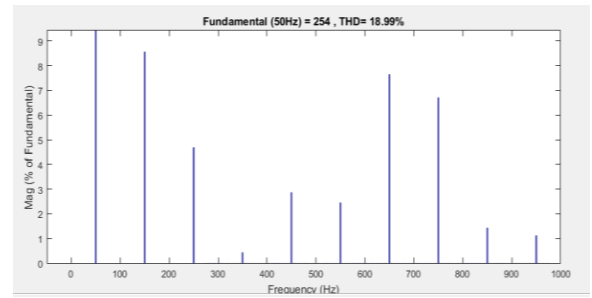

Fig. 20. THD values of the output voltage of the 7 level modified inverter with RL load

CONCLUSION

Upon comparing the output voltages obtained from both 5 and 7 level inverter, the output of the 7 level inverter is found to be closer to a sine wave and also has lower harmonic content than the 5 level inverter. The advantage of the modified configuration lies in the fact that it uses lesser number of switches than the conventional H bridge configuration. This results in lower costs, lower switching losses, and lower electromagnetic interference. Table 5 represents the THD results obtained for conventional 5 level & 7 level inverter configurations.

TABLE 5: THD values for 5 and 7 level conventional inverters

| Type of Conventional Multi level Inverter | THD with R Load | THD with RL Load |
|---|---|---|
| Conventional 5 level H Bridge inverter | 28.98 | 33.78 |
| Conventional 7 level H Bridge inverter | 16.72 | 18.38 |

Simulation results presented in table 5 reveals that the total harmonic distortion is reduced to 16.72 % from 28.98 % in 7 level conventional H Bridge configuration when compared with 5 level H bridge configuration.

TABLE 6: THD values for 5 level and 7 level modified inverters

| Type of Modified Multi level Inverter | THD with R Load | THD with RL Load |
|---|---|---|
| Modified 5 level inverter | 28.98 | 33.79 |
| Modified 7 level inverter | 17.02 | 18.99 |

Simulation results presented in table 6 reveals that the total harmonic distortion is reduced to 17.02 % from 28.98 % in 7 level modified H Bridge configuration when compared with 5 level modified bridge configuration. So it is concluded that modified 7 level H bridge configuration is better than modified and conventional 5 level inverters. Hence can be considered for the further studies.

REFERENCES

[1] Pallavi Appaso Arbune, Dr. Asha Gaikwad. "Comparative Study of Three level and Five level Inverter". IJAREEIE Vol. 5, Issue 2, 2016.

[2] Babaei, Ebrahim, Sara Laali, and Zahra Bayat. "A single-phase cascaded multilevel inverter based on a new basic unit with reduced number of power switches." *IEEE Transactions on industrial electronics* 62.2 (2015): 922-929.

[3] Nithya, S., C. R. Balamurugan, and S. P. Natarajan. "Design and analysis of new multilevel inverter topology with induction motor load." *Information Communication and Embedded Systems (ICICES), 2014 International Conference on*. IEEE, 2014.

[4] Lakshmi, T. V. V. S., et al. "Cascaded seven level inverter with reduced number of switches using level shifting PWM technique." *Power, Energy and Control (ICPEC), 2013 International Conference on*. IEEE,

[5] Nedumgatt, Jacob James, et al. "A multilevel inverter with reduced number of switches." *Electrical, Electronics and Computer Science (SCEECS), 2012 IEEE Students' Conference on*. IEEE, 2012.

[6] Pharne, I. D., and Y. N. Bhosale. "A review on multilevel inverter topology." *Power, Energy and Control (ICPEC), 2013 International Conference on*. IEEE, 2013.

[7] Ebrahimi, Javad, Ebrahim Babaei, and Gevorg B. Gharehpetian. "A new multilevel converter topology with reduced number of power electronic components." *IEEE Transactions on industrial electronics* 59.2 (2012): 655-667.

[8] Ebrahimi, Javad, Ebrahim Babaei, and Goverg B. Gharehpetian. "A new topology of cascaded multilevel converters with reduced number of components for high-voltage applications." *IEEE Transactions on power electronics* 26.11 (2011): 3109-3118.

[9] Najafi, Ehsan, and Abdul Halim Mohamed Yatim. "Design and implementation of a new multilevel inverter topology." *IEEE transactions on industrial electronics* 59.11 (2012): 4148-4154.

[10] Babaei, Ebrahim. "A cascade multilevel converter topology with reduced number of switches." *IEEE Transactions on power electronics* 23.6 (2008): 2657-2664.

[11] Rodriguez, Jose, Jih-Sheng Lai, and Fang Zheng Peng. "Multilevel inverters: a survey of topologies, controls, and applications." *IEEE Transactions on industrial electronics* 49.4 (2002): 724-738.

[12] Lezana, Pablo, José Rodríguez, and Diego A. Oyarzún. "Cascaded multilevel inverter with regeneration capability and reduced number of switches." *IEEE Transactions on Industrial Electronics* 55.3 (2008): 1059-1066.